\documentclass{aa}  

\usepackage{comment}
\usepackage{xcolor}
\usepackage{graphicx}
\usepackage{txfonts}
\usepackage{natbib}
\bibpunct{(}{)}{;}{a}{}{,}

\defcitealias{CS2017}{CS17}
\defcitealias{CS2020}{CS20}
\defcitealias{KC2016}{KC16}
\defcitealias{Gizon2020}{G20}

\begin{document}

\title{A Babcock-Leighton dynamo model of the Sun incorporating toroidal flux loss and the helioseismically-inferred meridional flow}

\titlerunning{A BL dynamo model incorporating toroidal flux loss and the observed meridional flow}
\authorrunning{S. Cloutier et al.}

\author{S. Cloutier$^1$,
          R. H. Cameron$^1$,
          \and
          L. Gizon$^{1,2}$
          }

\institute{Max-Planck-Institut f{\"u}r Sonnensystemforschung, Justus-von-Liebig-Weg 3, D-37077 G{\"o}ttingen, Germany\\
              \email{cloutier@mps.mpg.de} 
              \and
              Institut f\"ur Astrophysik, Georg-August-Universit\"at G\"ottingen, D-37077 G{\"o}ttingen, Germany}

\abstract
{Key elements of the Babcock-Leighton model for the solar dynamo are increasingly constrained by observations.}
{We investigate whether the Babcock-Leighton flux-transport dynamo model remains in agreement with observations if the meridional flow profile is taken from helioseismic inversions. Additionally, we investigate the effect of the loss of toroidal flux through the solar surface. 
}
{We employ the two-dimensional flux-transport Babcock-Leighton dynamo framework. We use the helioseismically-inferred meridional flow profile, and include toroidal flux loss in a way that is consistent with the amount of poloidal flux generated by Joy's law. Our model does not impose a preference for emergences at low latitudes, we do however require that the model produces such a preference.}
{We can find solutions in general agreement with observations, including the latitudinal migration of the butterfly wings and the cycle's 11~year period. The most important free parameters in the model are the depth to which the radial turbulent pumping  extends and the turbulent diffusivity in the lower half of the convection zone. We find that the pumping needs to extend to depths of about $0.80R_{\odot}$ and the bulk turbulent diffusivity needs to be around 10 km$^2$/s or less. We find that the emergences are restricted to low latitudes without the need to impose such a preference.
}  
{The flux-transport Babcock-Leighton model, incorporating the helioseismically inferred meridional flow and toroidal field loss term, is compatible with the properties of the observed butterfly diagram and with the observed toroidal loss rate. Reasonably tight constraints are placed on the remaining free parameters. The pumping needs to be to just below the  depth corresponding to the location where the meridional flow changes direction, and where numerical simulations suggest the convection zone becomes marginally subadiabatic. Our linear model does not however reproduce the observed ``rush to the poles'' of the diffuse surface radial field resulting from the decay of sunspots -- reproducing this might require the imposition of a preference for flux to emerge near the equator.} 

               
\maketitle

\section{Introduction} \label{sect:intro}
The solar cycle is driven by a self-excited fluid dynamo which is induced by the interaction between the large-scale magnetic field and flows within the convection zone of the Sun \citep{Ossendrijver2003,Charbonneau2014}. In the first part of the dynamo loop, differential rotation winds up poloidal magnetic field generating  toroidal field. This is the so-called $\Omega$-effect. The $\Omega$-effect is both well understood and constrained by observations. 

In the second part of this loop, the toroidal field generates new poloidal magnetic field. This new poloidal field has the opposite polarity to the original poloidal field. Each 11-year sunspot (or Schwabe) cycle is half of the 22-year magnetic (or Hale) cycle required to revert to the original polarity. Non-axisymmetric flows and fields play a critical role during this phase of the cycle. The non-axisymmetric processes involved in the second phase, however, are far from being either well understood or constrained. 

A major success of helioseismology was the determination of the sub-surface solar rotation profile, which challenged the dynamo-wave paradigm \citep{GT1991}. This challenge lead to the flux-transport dynamo (FTD) model \citep{Wang1991} where the deep meridional circulation causes the emergence locations of sunspots to drift equatorwards during a solar cycle. Observational and theoretical studies \citep[see eg.][]{DE2010,KO2011,CS2015} provide strong support for the Babcock-Leighton mechanism \citep[BL -][]{Babcock1961,Leighton1964,Leighton1969} to be the dominant mechanism in the second part of the dynamo loop, as opposed to the turbulent $\alpha$-effect \citep{Parker1955,Steenbeck1966}. At the core of the BL mechanism is the role played by the surface field. Sunspots emerge in bipolar magnetic pairs, with an east-west orientation usually in accordance with Hale's law. There is also a statistical tendency called Joy's law where the following spots to emerge closer to the poles and the leading spots to emerge closer to the equator. Sunspots decay within a few days to months, after which the field is dispersed by small-scale convective motions and transported poleward by the meridional flow. The flux cancellation of leading sunspot fields across the equator allows for the net buildup of a polar field by trailing sunspot fields.

Transport processes are required at the surface in order to transport the radial field from the equator to the poles, and to transport the subsurface toroidal field equatorwards to account for the equatorial migration of the butterfly wings (Sp\"orer's law). The surface part of the required transport has been established by observations of the surface meridional flow and the success of the surface flux transport model. The helioseismically inferred subsurface meridional flow is a relatively new constraint for these models. The use of the helioseismically-inferred meridional flow profile removes a number of free parameters from Babcock-Leighton type models. This makes a comparison with the observations a tighter test of the model and allows us to better constrain the remaining free parameters.

An additional recent constraint is that toroidal flux is lost through flux emergence, with a timescale estimated to be around 12 years by \citet[][hereafter \citetalias{CS2020}]{CS2020}. This paper will investigate if the BL FTD model, using the  meridional flow inferred by \citet[][hereafter \citetalias{Gizon2020}]{Gizon2020}, and including the toroidal flux loss associated with flux emergence, is consistent with observations. To this end we introduce a loss term in the evolution equation for the toroidal field that is consistent with the evolution of the poloidal flux associated with Joy's law. 

In the Babcock-Leighton type of model considered in this paper, the turbulent convective motions are not explicitly simulated, instead their affect on the magnetic field is parameterized \citep[eg.][]{Charbonneau2014}. Mean-field theory \citep{Moffatt1978,KR1980} actually shows that including the effect of turbulence introduces a large number of parameters. In the Babcock-Leighton model only a few are kept, the most important of which include the $\alpha$-effect, an increased turbulent diffusion, and downward diamagnetic pumping. Of these, the Babcock-Leighton $\alpha$-effect is poorly understood but well constrained by observations, while turbulent pumping and turbulent diffusion are largely unconstrained. 

In this paper we will see if the FTD model, with the observed meridional flow and flux loss, is compatible with the Babcock-Leighton model, and what constraints it places on the other processes of the model.

\section{Model}
\subsection{Dynamo equations}
In mean-field theory, the axisymmetric large-scale magnetic and velocity fields are decomposed into poloidal and toroidal components as:
\begin{equation}
\boldsymbol{B}=\nabla\times [A(r,\theta,t)\boldsymbol{\hat{e}}_{\phi}]+B(r,\theta,t)\boldsymbol{\hat{e}}_{\phi}, \label{vecb}
\end{equation}
\begin{equation}
\boldsymbol{u}=\boldsymbol{u}_m(r,\theta)+r\sin{\theta}\ \Omega(r,\theta)\boldsymbol{\hat{e}}_{\phi},
\end{equation}
where $A$ is the $\phi$-component of the vector potential field and $B$ is the toroidal component of the large-scale magnetic field, $\boldsymbol{u}_m$ is the meridional circulation, and $\Omega$ is the differential rotation. For an isotropic turbulent diffusivity that has only a $r$-dependence, the kinematic mean-field dynamo equations are:
\begin{equation}
\frac{\partial A}{\partial t}=-\frac{\boldsymbol{u}_p}{\varpi}\cdot\nabla (\varpi A)+\eta\left(\nabla^2-\frac{1}{\varpi^2}\right)A+S, \label{Aeq}
\end{equation}
\begin{equation}
\begin{aligned}
\frac{\partial B}{\partial t}=&-\varpi\boldsymbol{u}_p\cdot\nabla\left(\frac{B}{\varpi}\right)+\eta\left(\nabla^2-\frac{1}{\varpi^2}\right)B+\frac{1}{\varpi}\frac{\partial (\varpi B)}{\partial r}\frac{d\eta}{dr}\\&-B\nabla\cdot\boldsymbol{u}_p+\varpi [\nabla\times (A\boldsymbol{\hat{e}}_{\phi})]\cdot\nabla\Omega -L, \label{Beq}
\end{aligned}
\end{equation}
where $\varpi =r\sin{\theta}$, $\boldsymbol{u}_p=\boldsymbol{u}_m+\boldsymbol{\gamma}$, $\eta$ is the turbulent diffusivity, $\boldsymbol{\gamma}$ the turbulent pumping, $S$ the BL source term, and $L$ the toroidal field loss term due to flux emergence. The source and loss terms will be discussed in Section \ref{sect:femer}. 

\subsection{Differential rotation and meridional circulation}

\begin{figure}
\resizebox{\hsize}{!}{\includegraphics{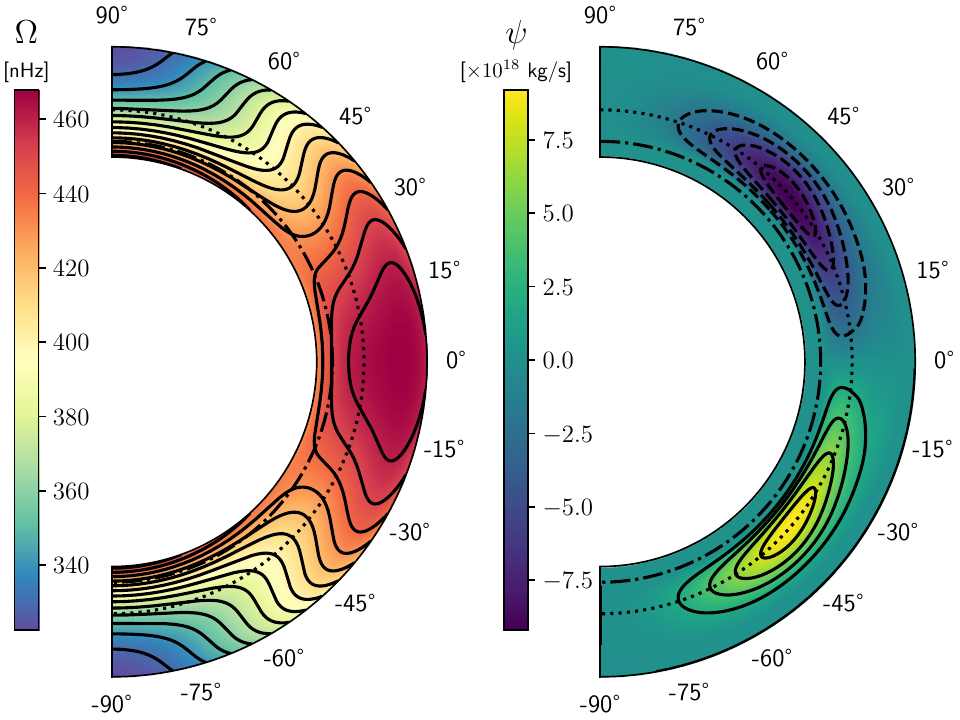}}
\caption{Rotation profile of \citet{Belvedere2000} given by equation \ref{diffrot} (left) and cycle-averaged and symmetrized stream function of the helioseismic meridional flow inversions of \citetalias{Gizon2020} (right). For the latter, positive values represent clockwise circulation and negative anticlockwise. The dash-dotted and dotted lines represent the approximate locations of the tachocline at $0.7R_{\odot}$ and reversal of the meridional flow direction at $0.8R_{\odot}$, respectively.}
\label{fig:flows}
\end{figure}

The large-scale flows need to be prescribed in kinematic models. For the differential rotation, we use the simple model provided by \citet{Belvedere2000} and shown in the left panel of Figure \ref{fig:flows}:
\begin{equation}
\Omega (r,\theta)=\sum_{j=0}^2\cos(2j\theta)\sum_{i=0}^4c_{ij}r^i, \label{diffrot}
\end{equation}
where the coefficients $c_{ij}$ can be found in that paper. This fit is an approximation of the helioseismologically inferred rotation rate of \citet{Schou1998}. 

As mentioned in the introduction, we will use the meridional circulation inferred from observations. In this study, we use the inversions of \citetalias{Gizon2020}. The authors furnish the meridional flow for cycles 23 and 24. In order to keep the parameter space study manageable, we take the average of the two cycles. In addition, since we are not here interested in the asymmetry between both hemispheres, we also symmetrize the flow across the equator. The right panel of Figure \ref{fig:flows} shows the meridional circulation we use in all our models. Note that the flow switches from poleward to equatorward at a radius of about 0.785$R_{\odot}$, which we will call the meridional flow turnover depth $r_t$, and the region beneath it the lower or deep convection zone, and the one above the upper or shallow convection zone. 

\subsection{Turbulent parameterizations}
We choose a  turbulent diffusivity profile which is written as a double step as in \citet{MJ2011}:
\begin{equation}
\begin{aligned}
\eta (r)=\eta_{\text{RZ}}+&\frac{\eta_{\text{CZ}}-\eta_{\text{RZ}}}{2}\left[1+\text{erf}\left(\frac{r-0.72R_\odot}{0.012R_\odot}\right)\right]\\
+&\frac{\eta_{R_\odot}-\eta_{\text{CZ}}-\eta_{\text{RZ}}}{2}\left[1+\text{erf}\left(\frac{r-0.95R_\odot}{0.01R_\odot}\right)\right],
\end{aligned}
\end{equation}
where $\eta_{\text{RZ}}=0.1$ km$^2$/s, $\eta_{R_\odot}=350$ km$^2$/s, and $\eta_{\text{CZ}}$ are respectively the radiative core, surface, and bulk values of the turbulent diffusivity. $\eta_{\text{CZ}}$ is a free parameter and $\eta_{R_\odot}$ has been chosen to be consistent with estimates from observations \citep[eg.][]{Komm1995}, the surface flux transport model of \citet{Lemerle2015}, and MLT \citep[eg.][]{MJ2011}. The values in the first error function have been chosen so that the drop in diffusivity occurs mostly before the helioseismically determined position of the tachocline \citep{Charbonneau1999}, roughly coinciding with the overshoot region \citep{CD2011}.

For the turbulent pumping, we adopt the profile given by \citet[][hereafter \citetalias{KC2016} -- see also the discussion in their section 2]{KC2016}:
\begin{equation} \label{eq:pump}
\boldsymbol{\gamma}=-\frac{\gamma_0}{2}\left[1+\text{erf}\left(\frac{r-r_{\gamma}}{0.01R_\odot}\right)\right]\boldsymbol{\hat{e}}_r,
\end{equation}
where here we take $r_{\gamma}=r_t=0.785R_{\odot}$. This choice will be discussed in Section \ref{sect:pump}.

\subsection{Flux emergence} \label{sect:femer}
The emergence of bipolar magnetic regions both removes toroidal flux \citepalias[see][]{CS2020} and creates poloidal flux (because of Joy's law). These two processes are clearly linked as are the respective loss and source terms in Equations~\ref{Beq} and \ref{Aeq}. 

We take the amount of flux emerging to be proportional to the toroidal flux density
\begin{equation}
b(\theta,t)=\int_{0.7R_\odot}^{R_\odot} B(r,\theta,t) r \mathrm{d}r,
\end{equation}
where the integration is over the depth of the convection zone. This prescription corresponds to a dynamo where the toroidal field is not necessarily stored near the tachocline but can be distributed throughout the convection zone \citep[\citetalias{KC2016},][]{Zhang2022}. It is in part motivated by observations of dynamos in fully convective stars \citep{WD2016}, and by cyclic dynamo action in 3D MHD simulations of spherical shells without a tachocline \citep[eg.][]{Brown2010,Nelson2013,Nelson2014}. The flux emergence rate $R$ can be written in general as a function of latitude:
\begin{equation}
R(\theta,t)=f_\theta(\theta) \frac{b(\theta,t)}{\tau_0},
\end{equation}
where $\tau_0$ is a timescale  and $f_\theta(\theta)$ is its latitudinal dependence, which we take to be
\begin{equation}
f_\theta(\theta)=\sin\theta,
\label{eq:f}
\end{equation}
corresponding to an emergence probability that is constant per unit length of the toroidal field lines.

In general the timescale, $\tau_0$ in Equation~9 depends on the dynamics associated with flux emergence. If these dynamics are dominated by the large-scale field than the buoyant rise time and $\tau_0$ to depend inversely on the mean-field value of $B^{2}$ \citep{KP1993}. If however small-scale magnetic fields remain coherent over timescales longer than the correlation time, then the $B$ filling factor which can be far from 1 becomes important. Some previous studies \citep[eg.][]{SS1989,Moss1990b,Moss1990a,JW1991} assume $\tau \sim B^{-2}$ so that the loss term scales like $B^3$. In this paper we consider the linear case where $\tau_0$ is a constant, and which would correspond to a case where the field is composed of filamentary structures with lifetimes longer than the turnover timescale of the turbulence and with local field strengths drawn from some distribution which is independent of flux. We stress that the aim in this paper is to consider a simple linear system. We defer the nonlinear case to future work.

The orientation of the flux emergence is governed by Joy's law which states that the leading polarity flux emerges on average closer to the equator than the trailing polarity one. We take the form of Joy's law used in \cite{Leighton1969},  $\sin{\delta}=\frac{1}{2} \cos\theta$, where $\delta$ is the angle between the solar equator and the line joining the two polarities. Then the rate at which toroidal flux density is lost due to flux emergence is
\begin{equation}
\begin{aligned}
\left.\frac{\partial b}{\partial t}\right|_{\mathrm{L}}
&= -\cos\delta~R(\theta,t),
 \\
  &= -f_\theta(\theta)\cos \delta \frac{b(\theta,t)}{\tau_0}, 
    \label{FL}  
\end{aligned}
\end{equation}
where the subscript $L$ indicates the contribution from the loss term. The tilting of a toroidal flux tube as it emerges gives rise to a $\theta$-component of the same polarity. The rate at which this $\theta$-component of the flux density is lost is
\begin{equation}
\begin{aligned}
\left.\frac{\partial}{\partial t}\int_{0.7 R_\odot}^{R_\odot} B_\theta(r,\theta,t) r\mathrm{d}r \right|_{\mathrm{S}}&= -\sin\delta~R(\theta,t), \label{eq:btheta}
 \\
  &= -f_\theta(\theta) \sin \delta\frac{b(\theta,t)}{\tau_0}.
\end{aligned}
\end{equation}
where the subscript $S$ indicates the contribution from the source term $S$. This is what gives rise to the Babcock-Leighton mechanism. From the definition of the poloidal field (Equation \ref{vecb}), 
\begin{equation}
B_\theta=-\frac{1}{r}\frac{\partial rA}{\partial r}.
\end{equation}
We choose a depth $R_b$ sufficiently below the base of the convection zone so that the 11-year cyclic component of the field is negligible there. Then multiplying both sides by $r$ and integrating from the base of the convection zone to the surface, we obtain 
\begin{equation}
\begin{aligned}
\int_{R_b}^{R_\odot}B_\theta(r,\theta,t)r\mathrm{d}r
=&-\left(A(R_\odot,\theta,t)R_{\odot}-A(R_b,\theta,t)R_b \right)  \\
=&-A(R_\odot,\theta,t)R_{\odot}.
\end{aligned}
\end{equation}
Therefore, in terms of the $\phi$-component of the poloidal vector potential $A$, Equation \ref{eq:btheta} becomes
\begin{equation}
\left.\frac{\partial A(R_\odot,\theta,t)}{\partial t}\right|_{\mathrm{S}}= f_\theta(\theta) \sin \delta\frac{b(\theta,t)/R_{\odot}}{\tau_0}.
\end{equation}

Next, we need to prescribe the radial structure of the source and loss terms. For the source term $S$, we follow \citetalias{KC2016} and assume
\begin{equation}
S(r,\theta,t)=
     f_r^S(r)\sin\theta\sin\delta\frac{b(\theta,t)/R_{\odot}}{\tau_0},
\end{equation}
where
\begin{equation}
f_r^S(r)=\frac{1}{2}\left[1+\text{erf}\left(\frac{r-r_S}{0.01R_\odot}\right)\right].
\end{equation}
$r_S$ is the depth to which the source extends. For most of the calculations, we choose $r_S=0.85R_{\odot}$ \citep[as in][]{MJ2010}. There are indications that this disconnection should happen much deeper than the usually assumed shallow location of $0.95R_{\odot}$ \citep{LC2002}. We will nevertheless vary this parameter in order to study its impact on the solutions. 

For the loss term $L$, we first distribute the toroidal flux density loss (equation \ref{FL}) over radius according to the amount of flux initially present, so that
\begin{equation}
L(r,\theta,t)=f_r^L(r) \sin\theta\cos\delta
\frac{B(r,\theta,t)}{\tau_0}, \label{l}
\end{equation}
where
\begin{equation}
f_r^L(r)=\frac{1}{2}\left[1+\text{erf}\left(\frac{r-0.70R_\odot}{0.01R_\odot}\right)\right].
\end{equation}

In order for the differential form of the source and loss terms to be valid, the emergence timescale $\tau_e$ must be considered infinitesimal with respect to the timescale over which the magnetic configuration of the large-scale field changes appreciably. It follows that a timescale separation must hold:
\begin{equation}
    \tau_e \ll P.
\end{equation}
In the case of the Sun, $\tau_e\sim 1$ day, and $P=11$~years, so the timescale separation is reasonable. This formulation implicitly ignores the effect of the meridional flow on the the emergence process. Nevertheless, this formulation is sufficient to study the general features of the solar cycle.

\subsection{Numerical procedure} \label{sect:num}
Equations \ref{Aeq} and \ref{Beq} are nondimensionalized and numerically solved in the meridional plane with $0\leq\theta\leq\pi$ and $0.65R_{\odot}\leq r\leq R_{\odot}$. We use a spatial resolution of $241\times 301$ evenly spaced in radius and colatitute grid points and a time step of $5\times 10^{-6}R_{\odot}^2/\eta_t$, where $\eta_t=10$ km$^2$/s. The inner boundary matches to a perfect conductor, so that:
\begin{equation}
A=0\quad\text{and}\quad\frac{\partial (rB)}{\partial r}=0\quad\text{at}\quad r=0.65R_{\odot},
\end{equation}
and the outer boundary condition is radial:
\begin{equation}
\frac{\partial (rA)}{\partial r}=0\quad\text{and}\quad B=0\quad\text{at}\quad r=R_{\odot}.
\end{equation}
The latter is necessary for FTD models to match surface flux transport models \citep{Cameron2012}. A second-order centered finite difference discretization is used for the spatial variables and the solution is forwarded in time with the ADI scheme \citep{Press1986}. We use the code initially developed by \mbox{D. Schmitt} in Göttingen \citep[as also used by][]{Cameron2012}. 

The linearity of equations \ref{Aeq} and \ref{Beq} allows us to choose $\tau_0$ and $\gamma_0$ such that the dynamo is approximately critical ($\sigma\leq 5\times 10^{-5}$ per year) with a cycle period of 12 years (within 0.1\%), roughly the average period of cycles 23 and 24. This way we reduce our parameter space to only one dimension ($\eta_{\text{CZ}}$). Our model being linear also means we can arbitrarily scale $A$ and $B$. In order to facilitate comparisons with observations, we scale the fields so that the maximum of the surface radial field is \mbox{10 G}, which is consistent with the observed polar field strengths at cycle minimum \citep[eg.][]{Hathaway2015}. 

\section{Observational constraints}
\label{sect:constraints}
\subsection{Toroidal flux loss timescale} 
The general expression for the toroidal flux decay timescale is
\begin{equation}
\tau (t)=-\frac{\Phi (t)}{\mathrm{d}\Phi (t)/\mathrm{d}t}, \label{taug}
\end{equation}
where $\Phi$ is understood as the net subsurface toroidal flux in the northern hemisphere:
\begin{equation}
    \Phi(t)=\int_0^{\pi/2}b(\theta,t)\mathrm{d}\theta,
\end{equation}
and $\frac{\mathrm{d}\Phi}{\mathrm{d}t}$ is its decay. 
In order to calculate the latter, we first need to specify the toroidal flux loss mechanism. For the loss $L$ due to flux emergence, we have
\begin{equation}
\left.\frac{\mathrm{d}\Phi}{\mathrm{d}t}\right|_{L}=-\int_0^{\pi/2}\int_{0.70 R_{\odot}}^{R_{\odot}}L(r,\theta,t)r\mathrm{d}r\mathrm{d}\theta, \label{eq:dphil}
\end{equation}
and its corresponding timescale will be denoted by $\tau_L$. Toroidal flux is also lost due to the explicit diffusion across the solar surface:
\begin{equation}
\left.\frac{\mathrm{d}\Phi}{\mathrm{d}t}\right|_\eta=\eta_{R_{\odot}}\int_0^{\pi/2}\left.\frac{\partial\left(r B\left(r,\theta,t\right)\right)}{\partial r}\right|_{R_{\odot}}\mathrm{d}\theta, \label{dphi}
\end{equation}
with an associated loss timescale of $\tau_{\eta}$. The observational constraint on the flux loss timescale is given by \citetalias{CS2020}, and corresponds to $\tau_L$ $\approx$ 12 years at solar maximum. Since these timescales vary across a cycle, we will calculate them at cycle maximum (to be defined in Section \ref{sect:pflux}).

It is possible to estimate the range of values $\tau_0$ should take. To do so we assume the tilt angle $\delta$ associated with Joy's law is small, so that $\cos\delta\sim 1$, and that the toroidal flux density can be approximated by $b(\theta,t)=b_0(t)\sin^m\theta$, where $b_0(t)$ is a time-dependent scalar, and $m$ determines how closely the field is concentrated near the equator. With these approximations, we obtain:

\begin{equation}
    1< \frac{\tau_L}{\tau_0} < \frac{\pi}{2},  
\end{equation}
where $\tau_0$ is the timescale for flux loss in the model, defined in equation 9, and the limits correspond to $m=0$ and $m=\infty$. The model parameter $\tau_0$ should thus be comparable to (not more than a factor of 2 smaller than) the observed toroidal flux loss timescale $\tau_L$.
 
\subsection{Polar cap and activity belt flux densities} \label{sect:pflux}
An important observational constrain is that the maxima of azimuthally averaged polar flux densities should be around the same strength as maximum flux densities in the activity belt. Since the evolution equations we are using are linear, we have nominally set the  maximum of the polar field to be 10 G. This implies that the azimuthaly averaged radial field in the butterfly wings in the model should also be around 10~G. This is a constraint on the model which we will return to when evaluating whether the model can produce solar-like cycles.

\subsection{Cycle phase of polar maxima} 
An important constraint we will take into account is when the maximum of polar flux occurs. As it is observed to happen quite close to the activity minimum, its corresponding phase shift of about $90^\circ$ with respect to cycle maximum. To measure this shift in the simulations, we need a definition for when the cycle maxima and the maxima of polar fields occur. 

We begin with the surface radial flux of the polar cap
\begin{equation}
\Phi_p(t) \equiv 2\pi R_{\odot}^2\int_{60\text{°}}^{90\text{°}}B_r(R_{\odot},\lambda,t)\, \mathrm{d}(\sin \lambda), \label{eq:phip}
\end{equation} 
and of the activity belt
\begin{equation}
\Phi_a(t) \equiv 2\pi R_{\odot}^2\int_{0\text{°}}^{40^\circ} B_r(R_{\odot},\lambda,t)\, \mathrm{d}(\sin \lambda). \label{eq:phia}
\end{equation} 
We then define cycle maximum times, $T_a$, as the times when the activity belt flux $\Phi_a$ is maximum. The times of the maximum polar flux $T_p$ is similarly defined. $T_a$ and $T_p$ are both defined by the signed fluxes, and hence each have one maxima per magnetic cycle (of about 24 years).

For each cycle, $i$, we then calculate the phase shift between the polar maximum times and the activity maximum times $\Delta\phi$ by
\begin{equation}
\Delta\phi=\pi [T_p(i+1)-T_a(i)] /P-\pi,
\end{equation}
where $P$ is the (activity) cycle period.

\section{Results \label{sect:results}}

We first present our reference model including the Babcock-Leighton loss term in Section \ref{sect:ref}, then examine parameter sensitivity (Sections \ref{sect:param} to \ref{sec:nolabel}), and finally gauge the importance of the dynamo wave (Section \ref{sect:dynwav}).

\subsection{Reference model} \label{sect:ref}

Our reference model has a bulk diffusivity of $\eta_{\text{CZ}}=10$ km$^2$/s, which places the simulation in the advection-dominated regime (an explanation of why such a low diffusivity is required in our setup is given in Section \ref{sect:diff}). Values of $\tau_0=12.5$ years and $\gamma_0=43.8$ m/s were required to achieve a critical 12-year (activity) cycle period dynamo. The reference model is presented in Figure \ref{fig:n1}. It can be seen that we are able to produce a reasonably solar-like butterfly diagram when using the helioseismically-inferred meridional flow of \citetalias{Gizon2020}.

\subsubsection{Comparison with observational constraints}
This model has an emergence loss timescale of $\tau_L=17.2$ years, somewhat longer than the 12 years inferred by \citetalias{CS2020}. The diffusive loss timescale, on the other hand, is $\tau_{\eta}=96.8$~years. The estimate of \citetalias{CS2020} is based on all toroidal flux escaping through the photosphere. The combined rate at which flux is lost through the photosphere in the model can be estimated to be $1/(1/\tau_L+1/\tau_{\eta})=14.6$~years and hence close to the inferred 12-year timescale. The value of $\Delta\phi$ from the model is noticeably larger than the observed value at $\Delta\phi=134$°. The maximum value of the surface radial field in the butterfly wings is around 5 G, so about half of the maximum polar field strength \citep[The observed average polar field is similar to the average in the butterfly wings, see e.g. the butterfly diagrams in][]{Hathaway2015}. Considering the polar field strength is somewhat uncertain, our results are not inconsistent with observations. Our simulations do not have the problem of very large polar fields typical of FTD models \citep{Charbonneau2020}.   

The net toroidal flux $\Phi$ (lower panel of Figure \ref{fig:fieldn1}) shows that it reaches a maximum value of about $5\times 10^{23}$ Mx close to cycle maximum, which is in rather good agreement with the estimates of \citet{CS2015} for cycles 22 and 23. 

\begin{figure}
\resizebox{\hsize}{!}{\includegraphics{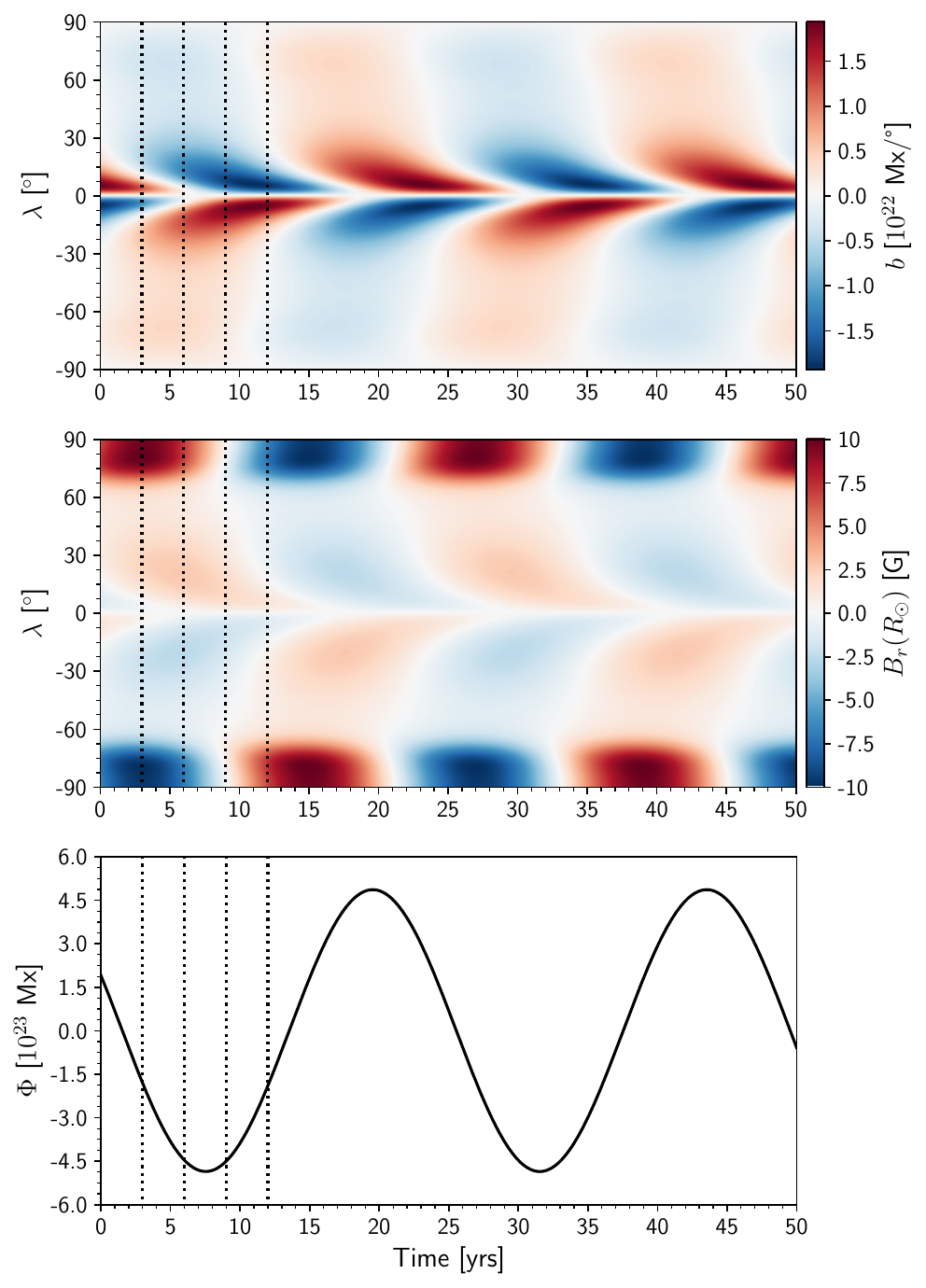}}
\caption{Time-latitude diagrams of the toroidal flux density $b$ (top), surface radial field $B_r(R_{\odot})$ (middle), and the net toroidal flux in the northern hemisphere $\Phi$ (bottom) for our reference model. The vertical dotted lines indicate the times where the snapshots of Figure \ref{fig:fieldn1} were taken.}
\label{fig:n1}
\end{figure}

\begin{figure}
\resizebox{\hsize}{!}{\includegraphics{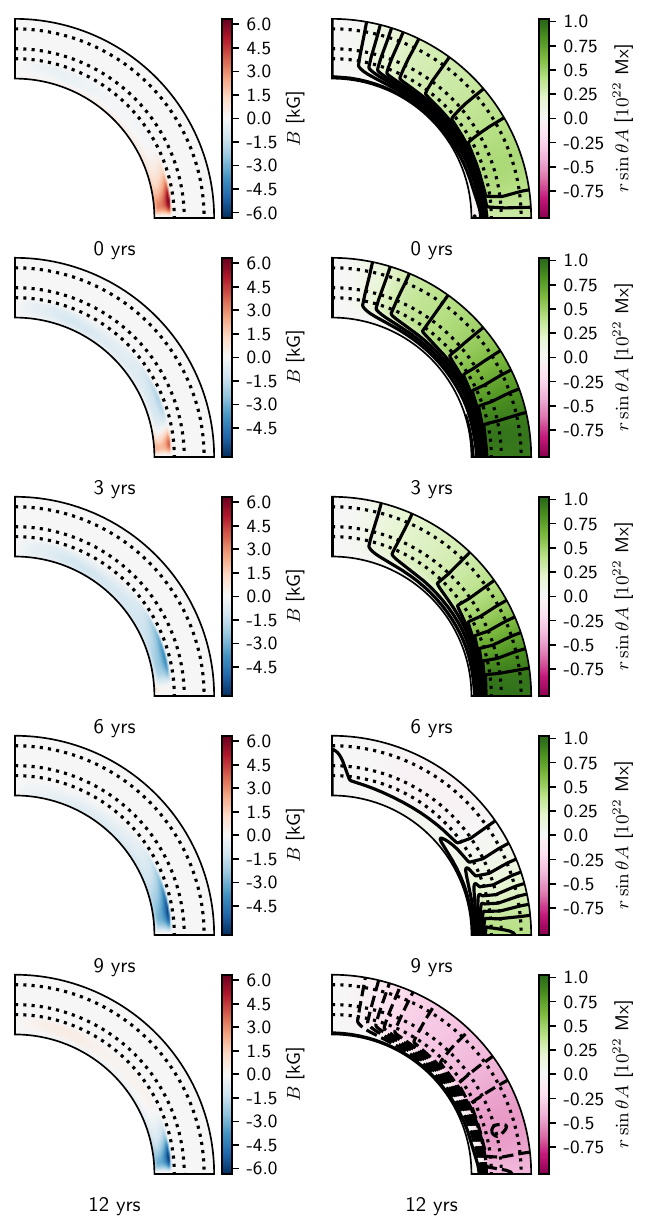}}
\caption{Meridional cuts of the North hemisphere toroidal field (left column) and poloidal field (as $r\sin\theta A$, right column) of the reference model for specific times indicated by the vertical dotted lines in Figure \ref{fig:n1}. The dotted lines are located at radii of $0.95$, $0.85$, and $0.80 R_{\odot}$, the approximate locations of the bottom of the near-surface shear layer, $r_S$, and $r_{\gamma}$ respectively.}
\label{fig:fieldn1}
\end{figure}

An important result of our model is that it achieves confinement of emergences to the low observed latitudes without the need for an emergence probability decreasing faster than $\sin\theta$ with latitude. This can be seen in the upper panel of Figure \ref{fig:n1}, where we see that  the toroidal flux density is mainly strong near the equator. The left panel of Figure \ref{fig:fieldn1} shows the toroidal field is mainly stored deep in the convection zone. This confinement of the toroidal field to deep in the convection zone is a consequence of the imposed radial pumping. The confinement to near the equator is then due to the meridional flow which advects the material from high latitudes towards the equator. The combination of radial pumping and equatorward meridional flow in the lower half of the convection zone leads to a stagnation point near the equator in the lower half of the convection zone where the field builds up until it is removed through emergence \citep[also see][]{CS2017,Jiang2013}.

Our simulated butterfly diagram differs from the observed one in that it lacks a distinct "rush to the poles" of the trailing diffuse field of the decayed sunspots.

\subsection{Parameter dependence} \label{sect:param}
Our model has 5 free parameters: the source and loss terms timescale $\tau_0$ and the depth where sunspots are anchored $r_S$, the turbulent pumping amplitude $\gamma_0$ and the depth it reaches down to $r_\gamma$, and the turbulent diffusivity in the bulk of the convection zone $\eta_{CZ}$. In this section we will first provide a qualitative description of what different choices of the parameters produce. 

Importantly, the results and constraints we find are for under the assumption that $f_\theta(\theta)=\sin\theta$, i.e. that there is no imposed preference for emergences to occur at low latitudes. We also performed simulations with $f_{\theta}(\theta)=\sin^{12}\theta$ \citepalias[as in][]{KC2016}, and as expected were are able to find critical dynamo solutions which match the observations for a much wider range of parameters. 

\subsubsection{Influence of the source depth} \label{sect:anchor}

We here investigate how the choice of $r_S$ affects the solutions. We used the same value of $\eta_{\text{CZ}}$ as in reference case, and varied $r_S$. The values of $\tau_0$ and $\gamma_0$ were then chosen so that the growth rate is zero and the cycle period is 12 years. We found that the solutions with flux loss are not very dependent on $r_S$ for $r_S$ extending from just above 0.78 to the surface. This is because the solutions with flux loss require strong pumping which rapidly stretches the poloidal field so that they extend radially to the depth at which the pumping stops. This makes the model insensitive to the initial depth of the poloidal source term in (at least in the region of parameter space near the reference case).

\subsubsection{The bulk diffusivity} \label{sect:diff}

We only find critical 12-year periodic solutions when the bulk diffusivity is of the order of $10$~km$^2$/s. This is a consequence of the strong radial shear of the equatorward component of the helioseismically-inferred meridional flow $u_{\theta}$ in the lower third of the convection zone. The strong radial shear leads to toroidal flux at different depths being advected in latitude at very different rates. This implies that flux originally concentrated at one latitude over a range of depths will be quickly spread out in latitude. 

To understand the role of the radial shear of the latitudinal flow, we can imagine toroidal field initially at one latitude but spread out in radius from  $r=0.766R_{\odot}$ to $0.785R_{\odot}$. These depths were chosen so that the meridional flow will vary from almost 1~m/s equatorwards to almost 0~m/s. Over 5 years this will spread the flux over a latitudinal band of $157$~Mm. This spreading out would be similar to a diffusivity of $(157)^2/5$~Mm$^2$/year $=156$~km$^2$/s. In the context of Babcock-Leighton FTD models, this is a large value. As a comparison, the 1D model of \citet{CS2017} which also assumes $f_\theta(\theta)=\sin\theta$, requires the latitudinal diffusivity in the bulk of the convection zone to be lower than about 100 km$^2$/s. This effective diffusivity is, however, in agreement with the estimate of of $150-450~$km$^2$/s of \citet{CS2016} based on the properties of the declining phase of solar cycle.

The essential point of the above is that, unless the toroidal field is confined to a narrow range of depths,
the latitudinal shear in the meridional velocity quickly spreads the toroidal field out in latitude. Consequently, if there is no imposed preference for emerging near the equator then the butterfly diagram ceases to be solar-like. The requirement for confinement in latitude is what imposes the constraint that $\eta_{CZ}\approx 10$~km$^2$/s. We consider this fixed for the rest of this paper. We also comment that if the radial shear in the differential rotation was weaker, then this constraint would be much weaker.

\subsubsection{Turbulent magnetic pumping} \label{sect:pump}
With our chosen $f_\theta(\theta)=\sin\theta$
(Eq.~\ref{eq:f}), we find growing dynamo solutions only for values of $r_{\gamma}$ not far away from $0.785R_{\odot}$. This depth is where the helioseismically-inferred meridional flow profile changes direction from poleward to equatorward, and roughly where numerical simulations suggest the convection zone might be weakly subadiabatic \citep{Hotta2017}. A slightly broader range of pumping depths can be achieved when the loss term is not included, but shallower depths make the butterfly wings broader. 

Our conclusion from this is that the pumping depth is fairly tightly constrained if the appearance of spots to low latitudes is only caused by the equatorward meridional flow leading to a build up at low latitudes. The depth of the pumping is poorly constrained if the preference for low latitude emergence is imposed.

Observations do not provide estimates for the amplitude of turbulent pumping at depth, and so it is interesting to compare our results with those from global MHD simulations. \citet{Shimada2022} find that in the outer half of the convection zone has a turbulent diffusivity of $\sim 10$ km$^2$/s, similar to the values we needed in our model without an imposed preference for emergence at low latitudes, they also find that $\gamma_r$ peaks at $\sim$ 10 m/s, while \citet{Simard2016} and \citet{Warnecke2018} find amplitudes of the order of $1-2$ m/s or half the root-mean-square velocity. The latter is of the order of 40 m/s according to mixing-length \citep{Vitense1953,BV1958} estimates. It thus appears the pumping velocities required in the reference case are too large by a factor of 2 to 3. We defer a discussion of this to Sections \ref{sect:inmer} and \ref{sec:nolabel}.

\subsection{Role of the toroidal flux loss term $L$} \label{sect:wloss}

\begin{figure}
\resizebox{\hsize}{!}{\includegraphics{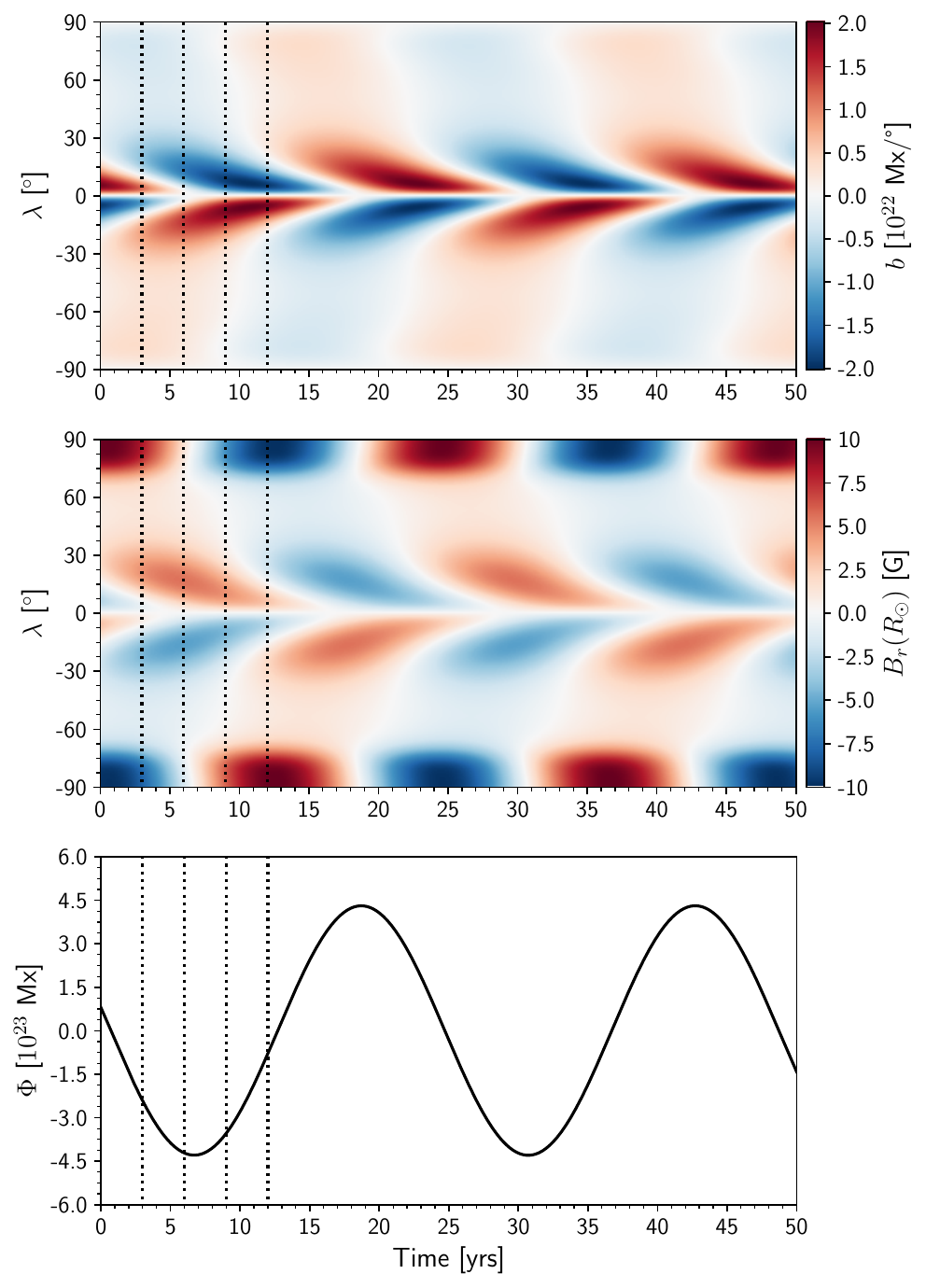}}
\caption{Time-latitude diagrams of the toroidal flux density $b$ (top), surface radial field $B_r(R_{\odot})$ (middle), and the net toroidal flux in the northern hemisphere $\Phi$ (bottom) for the case without the flux loss associated with flux emergence (and with a period of 12 years and zero growth rate).  
The vertical dotted lines indicate the times where the snapshots of Figure \ref{fig:fieldn1w} were taken.}
\label{fig:n1w}
\end{figure}

\begin{figure}
\resizebox{\hsize}{!}{\includegraphics{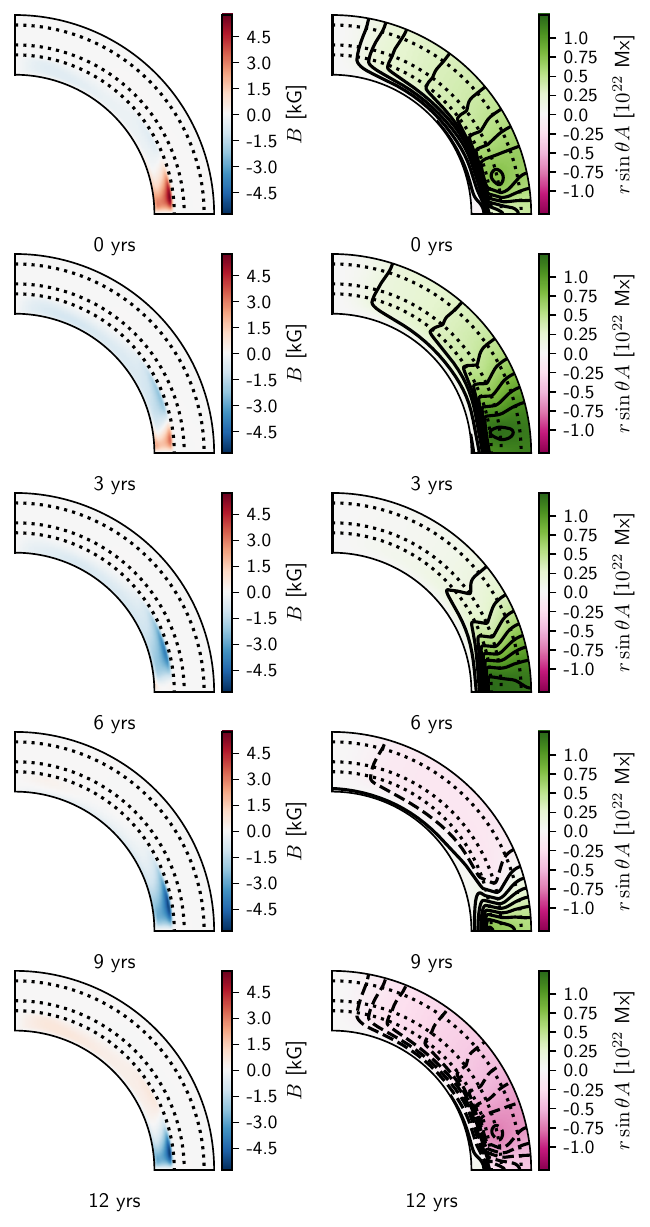}}
\caption{Meridional cuts of the North hemisphere toroidal field (left column) and poloidal field (as $r\sin\theta A$, right column) of the the case without the flux loss associated with flux emergence (and with a period of 12 years and zero growth rate) for specific times indicated by the vertical dotted lines in Figure \ref{fig:n1w}. The dotted lines are located at radii of $0.95$, $0.85$, and $0.80 R_{\odot}$, the approximate locations of the bottom of the near-surface shear layer, $r_S$, and $r_{\gamma}$ respectively.}
\label{fig:fieldn1w}
\end{figure}

In order to gauge the importance of emergence loss term, in this subsection we switch off the term by setting $L=0$ in Equation~4. We first use the same parameters as in the reference case. The resulting cycle period is shorter at 11.3 years. However, the solution is rapidly growing, with a growth rate of  about 66\% per cycle. This growth is not unexpected, as emergences are no longer able to remove the subsurface toroidal flux and it must now be removed either through its "unwinding" by the new cycle flux, or by diffusive cancellation across the equator. Because emergences no longer deplete the subsurface toroidal flux, more poloidal field is generated so that the polar fields are reversed much faster, explaining the shorter period.

We also investigated the 12~year period critical solutions when $L=0$. Doing so required $\tau_0=9.3$~years (as against the reference case where $\tau_0=12.5$~years), and a  turbulent pumping  $\gamma_0=6.41$~m/s. The time-latitude diagrams and meridional cuts for this case are shown in figures \ref{fig:n1w} and \ref{fig:fieldn1w}. The surface diffusion loss timescale is reduced to $\tau_{\eta}=25.6$ years, only a factor of two larger than the observed value of the toroidal flux loss timescale. This is due to the lowered pumping amplitude, which makes the diffusion of the toroidal field through the surface less difficult than in the case with strong pumping. Even with such low pumping, the poloidal field near the surface is still almost radial (Figure \ref{fig:fieldn1w}).

Looking at the butterfly diagram, the most apparent difference is the large decrease of the polar field strengths compared to the fields in the butterfly wings. The maximum value of the latter goes up from about 5 to 7.5 G, which is relatively close to the observed value of around 10 G. In this case, we find maxima phase between the polar field maxima and active region maxima is $101^{\circ}$, similar to that which is observed. 
 
Note that the pumping amplitude of $\gamma_0=6.41$~m/s in this model is in much better agreement with estimates from global MHD simulations (cf. Section \ref{sect:pump}). This is because models without the loss term achieve shorter periods much more easily. In principle, then, very large pumping velocities are not necessary to obtain a functioning dynamo for this class of models.

\subsection{Sensitivity to the meridional flow and differential rotation} \label{sect:inmer}

Here we investigate how sensitive the simulations are to the meridional flow and differential rotation. We do this by considering the inferred meridional flow for cycles 23 and 24 separately, and a differential rotation profile which differs significantly from the one of the reference model at high latitudes. As is also the case for the reference solution, we do not impose a preference for emergences at low latitudes (if we impose a preference for emergences at low latitudes, then the parameter space where the model has similar properties to the observations becomes much larger). As all solutions mentioned in this section have qualitatively the same butterfly diagrams as the reference case they are not shown.

First, we consider individually the symmetrized (across the equator) meridional flow profiles of cycles 23 and 24. Using the meridional flow from cycle 23, a 12-year periodic critical dynamo requires $\gamma_0=16.3$ m/s. This is a substantial reduction from the reference case. Increasing the period to 13.3 years and keeping the criticality requirement led to pumping speeds of $\gamma_0=10$ m/s. Using the meridional flow from cycle 24, we were unable to find critical solutions with periods shorter than 12 years. A critical solution with $\gamma_0=10$ m/s required a period of 16.6 years.

Clearly the model, where emergence is not restricted to low latitudes, is very sensitive to the meridional flow. This is because emergences at high latitudes are inefficient at getting flux across the equator, which is what eventually reverses the polar fields. The observational constraint, that the cycle period is similar to the timescale for which toroidal field is lost through the surface due to flux emergence, implies that the cycle period involves a balance between the flux transport to low latitudes and the loss
through emergence.

In this context, both mean-field theory \citep[eg.][]{Kitchatinov1991,KN2016} and global  numerical models \citep[eg.][and references therein]{Shimada2022} indicate equatorial latitudinal turbulent pumping could also be substantial. From the BL-FTD modelling this would correspond to an increase in the meridional return flow, and would lead to a reduction in the strength of the required radial pumping.

Second, we consider the sensitivity to differential rotation. Again we consider critical 12-year periodic solutions, using the average meridional flow profile used in the reference case. We now use the differential rotation profile of \citet{LS2018}, which differs from that of the reference at high latitudes (rather than the analytic fit of \citet{Belvedere2000} often used in dynamo studies and used in the reference case). The required parameter values are $\tau_0=11.8$ yrs and $\gamma_0=28.5$ m/s. $\tau_0$ is again of the order of cycle period and the 12-year estimate for the toroidal flux loss timescale of \citet{CS2020}. But the pumping velocity is much more reasonable. 

\subsection{Sensitivity to assumption that growth rate is zero and period is 12 years}
\label{sec:nolabel}
\begin{figure}
\resizebox{\hsize}{!}{\includegraphics{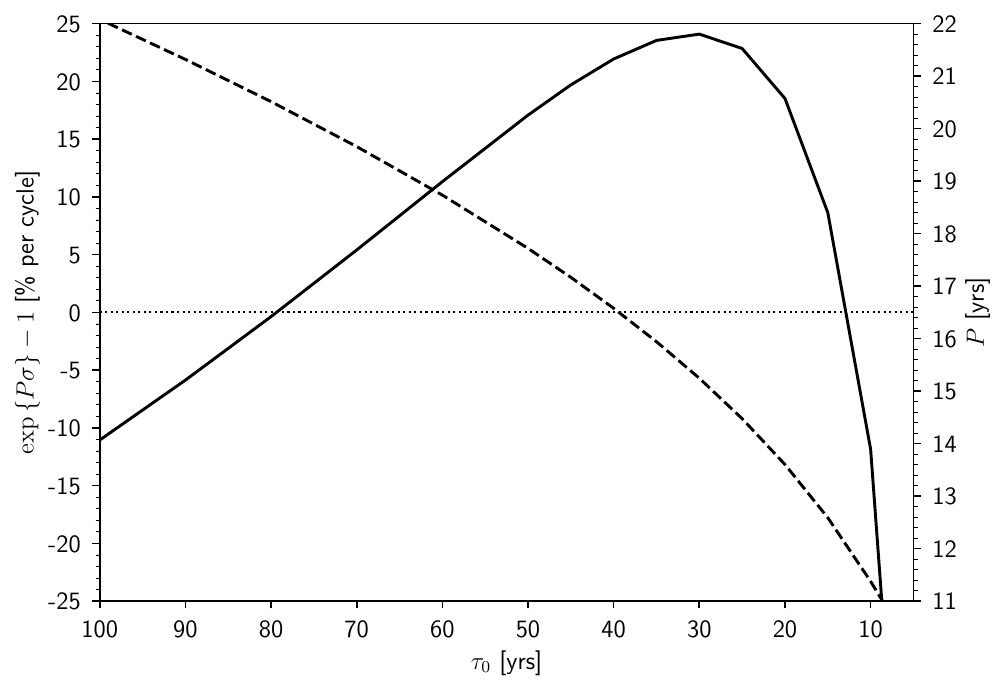}}
\caption{Percental per-cycle growth of the toroidal flux (solid line, left axis) and cycle period (dashed line, right axis) as a function of the timescale parameter $\tau_0$.}
\label{fig:tau0}
\end{figure}

We have thus far concentrated on the kinematic regime with zero growth rates. The Sun is certainly in a statistically saturated state. The kinematic case with growth rate zero is relevant if the system is weakly nonlinear. Whether or not this is the case for the Sun is open \citep[for arguments in favour of this see][]{vanSaders2016,Metcalfe2016,KO2017}. In the strongly nonlinear case, the period will be substantially affected by the choice of the nonlinearity and the growth rate in the linear regime is no longer a constraint. The observations are thus less constraining in the strongly nonlinear case. For this reason, we have focused on the weakly nonlinear case and have looked for zero-growth rate solutions to the linear problem. The addition of a weak nonlinearity will slightly modify both the growth rate (in the saturated state it will be zero) and period. Hence in this section we consider the sensitivity of the growth rate and periods to $\tau_0$ and $\gamma_0$.

Figure \ref{fig:tau0} shows the cycle period and the growth of the toroidal flux per cycle as a function of the timescale parameter $\tau_0$. As in \citetalias{KC2016}, we observe that increasing the source term amplitude $\tau_0^{-1}$ causes the growth rate to increase until the cycle period becomes too short for the meridional flow to transport the field (see Section 4.1 of \citetalias{KC2016}). Eventually, the dynamo shuts down completely. Growing solutions can nonetheless be reached by further increasing the source term amplitude $\tau_0^{-1}$. However, the resulting cycle periods are very short ($\lesssim 3$ years) and the dynamo is now driven by a dynamo wave propagating equatorwards in the high-latitude tachocline.

\begin{figure}
\resizebox{\hsize}{!}{\includegraphics{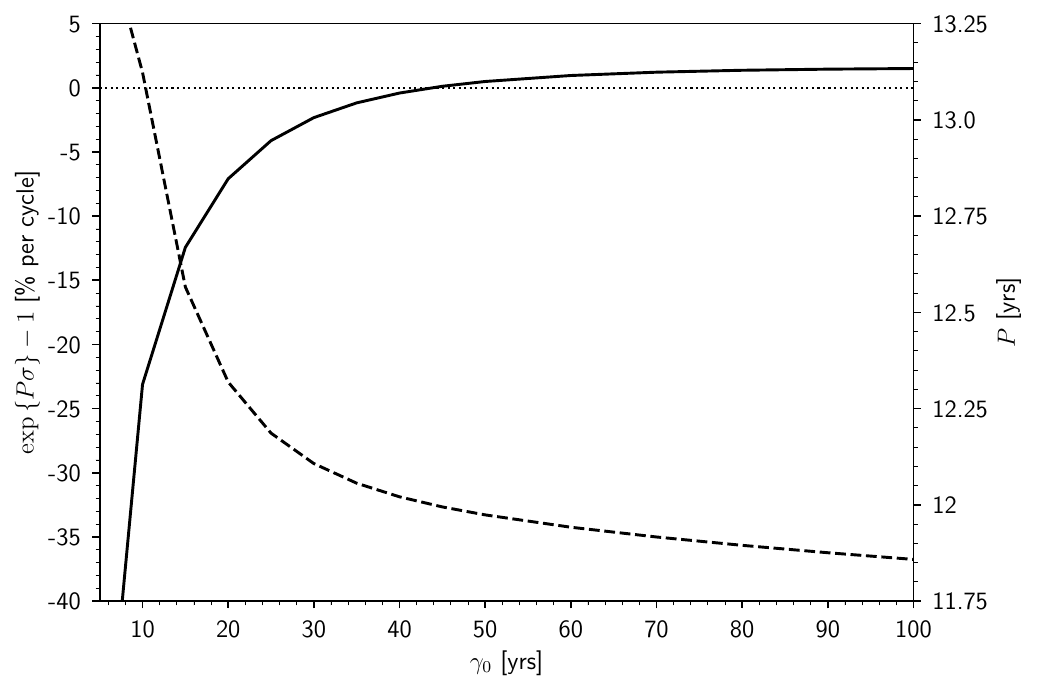}}
\caption{Percental per-cycle growth of the toroidal flux (solid line, left axis) and cycle period (dashed line, right axis) as a function of the pumping amplitude $\gamma_0$.}
\label{fig:gam0}
\end{figure}

The effect of the pumping amplitude on the growth rate and cycle period is shown in Figure \ref{fig:gam0}. The growth rate is very sensitive to the pumping amplitude at lower values, where the operating threshold is not yet met, as flux emergence then quickly removes the toroidal flux at high latitudes. But the effect of pumping saturates as its amplitude increases. At some point the time required for the poloidal field to reach the lower convection zone is essentially instantaneous. Note that we have concentrated on solutions near the bifurcation point where dynamo action switches on. This very likely makes the model 
more sensitive to the different parameters than would be the case if we were considering a non-linear, saturated dynamo.

\subsection{Role of the dynamo wave} \label{sect:dynwav}

\begin{figure}
\resizebox{\hsize}{!}{\includegraphics{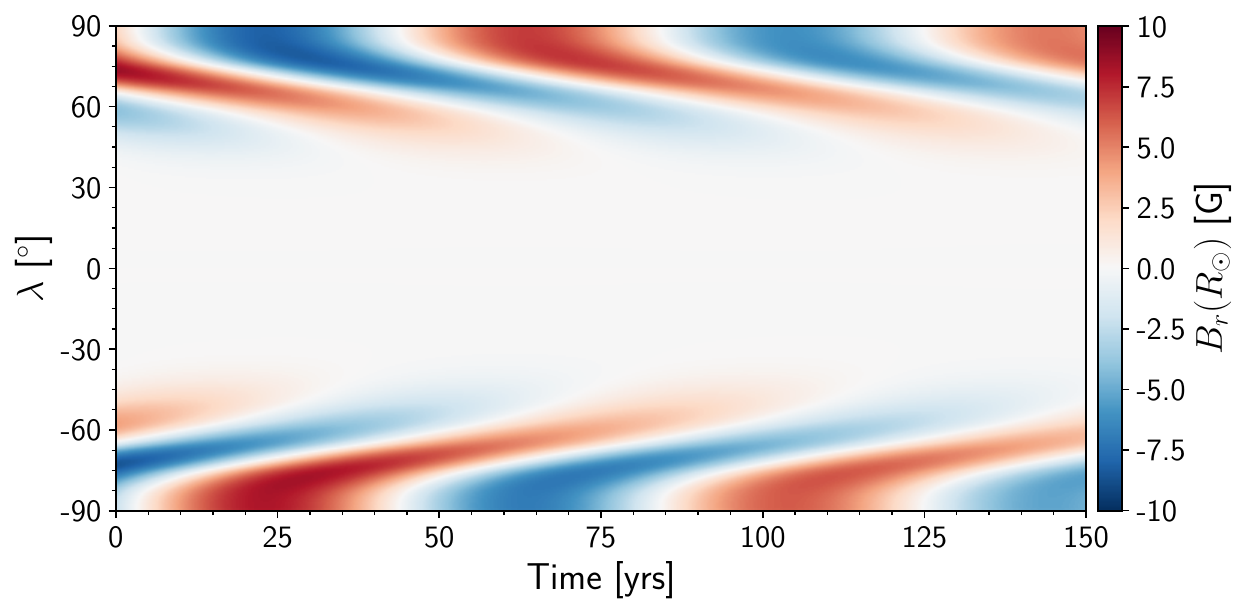}}
\caption{Butterfly diagram of our reference model where the equatorward component of the meridional flow was shut off.}
\label{fig:n1mfo}
\end{figure}

To investigate what role the subsurface meridional return flow is playing, we apply the same procedure as \citetalias{KC2016} to our reference model, namely we switch off the equatorward component of the meridional flow (see Section 3.2 of \citetalias{KC2016} for a discussion). Figure \ref{fig:n1mfo} shows the resulting magnetic field butterfly diagram. For the reference parameters this mode is decaying (and so is not a dynamo) with fields almost entirely located above $45^{\circ}$. Since there is no equatorward component of the meridional flow, the equatorward migration of the field is due to the negative radial rotation shear in the high-latitude tachocline, and the direction of propagation is in accordance with the Parker-Yoshimura sign rule. We hence, not surprisingly, conclude that the subsurface meridional flow is essential in this model.

\section{Conclusion}

Using the helioseismically-inferred meridional flow of \citetalias{Gizon2020}, we have shown that the Babcock-Leighton FTD model remains generally consistent with observations. We have also shown that the long-standing problem of the latitudinal distribution of sunspots can be solved if turbulent pumping reaches depths just under $0.80R_{\odot}$, but not much deeper, where the meridional flow's direction switches from poleward to equatorward. High turbulent pumping velocities are necessary to essentially store the toroidal flux under this location, in agreement with the results of \citetalias{Gizon2020} \citep[see also][]{Parker1987}. There, the meridional flow, in conjunction with the $\Omega$-effect through the latitudinal shear present in the bulk of the convection zone (which is maximal at mid-latitudes), causes an accumulation of toroidal flux at equatorial latitudes. Turbulent pumping effectively short-circuits the meridional circulation, preventing significant generation of toroidal field at high latitudes. No additional restriction of  emergences to low latitudes is required.  

Our model using the helioseismically inferred meridional flow, and including the observed toroidal flux loss associated with flux emergence in a way that is consistent with the Babcock-Leighton source term, is able to reproduce the observed properties of the solar cycle, including the latitudinal migration of the sunspot wings and the approximately 11~year period. Our reference model predicts a toroidal flux loss timescale of 14.8 years at cycle maximum, compared to the estimate of 12 years of \citetalias{CS2020}.

\begin{acknowledgements}
    The authors wish to thank the anonymous referee for comments that helped improve the overall quality of this paper.
      SC is a member of the International Max Planck Research
School for Solar System Science at the University of Göttingen. The authors acknowledge partial support from ERC Synergy
grant WHOLE SUN 810218.
\end{acknowledgements}

\bibliographystyle{aa}
\bibliography{cloutier2023}

\end{document}